\documentclass[aps,twocolumn,preprintnumbers,showpacs,amsmath,nofootinbib,secnumarabic,floatfix]{revtex4}
\def\scn#1#2{\section{#1}\lb{#2}}
\def\sscn#1#2{\subsection{#1}\lb{#2}}

\usepackage{epsfig,amssymb,verbatim}

\def\bfl{\begin{flushleft}}
\def\efl{\end{flushleft}}
\def\bfr{\begin{flushright}}
\def\efr{\end{flushright}}
\def\bc{\begin{center}}
\def\ec{\end{center}}
\def\be{\begin{equation}}
\def\ee{\end{equation}}
\def\ba{\begin{eqnarray}}
\def\ea{\end{eqnarray}}
\def\baa#1{\begin{array}{#1}}
\def\eaa{\end{array}}
\def\bw{\begin{widetext}}
\def\ew{\end{widetext}}
\def\nn{\nonumber }
\def\lb#1{\label{#1}}
\def\bit{\begin{itemize}}
\def\eit{\end{itemize}}
\def\bcm{}
\newcommand{\comments}[1]{}

\def\schrod{Schr\"odinger}

\newtheorem{Def}{Def.}[section]

\newtheorem{Prp}[Def]{Proposition}

\begin{document}

\preprint{\small Eur. Phys. J. B 85 (2012) 273
[arXiv:1204.4652]}

\title{
Volume element structure and  roton-maxon-phonon excitations
in superfluid helium beyond the Gross-Pitaevskii approximation
}

\author{Konstantin G. Zloshchastiev}

\affiliation{
School of Chemistry and Physics, 
University of KwaZulu-Natal
in Pietermaritzburg,
Private Bag X01, 
Scottsville 3209,
Pietermaritzburg, 
South Africa}


\begin{abstract}
We propose a theory which deals with the structure and interactions of volume elements
in liquid helium II.  
The approach consists of two nested models linked via parametric space.
The short-wavelength part describes the interior structure of the fluid element  
using a non-perturbative approach based on the logarithmic wave equation; it suggests
the Gaussian-like behaviour of the element's interior density  
and interparticle interaction potential. 
The long-wavelength part is the quantum many-body theory of such elements
which deals with their dynamics and interactions. 
Our approach leads to a unified description of the phonon, maxon and roton excitations,
and has noteworthy agreement with experiment: 
with one essential parameter to fit we reproduce at high accuracy not only the roton minimum but also the neighboring local maximum as well as the sound velocity and structure factor.
\end{abstract}


\maketitle

\scn{Introduction}{s-intro}

The microscopical structure of quantum liquids
is different
from
that of classical ones. The latter are discrete fluids 
consisting
of localized atoms or molecules which thermal de Broglie length
is smaller than the average atomic or molecular separation.
For instance, the thermal
de Broglie wavelength of a water molecule at room temperature
is less than the radius
of the hydrogen atom. 
Therefore,
classical liquids can be treated as continuous media only on the length scales
which are larger than  atomic or molecular separation.
In  quantum liquids it is other way around:
atoms become delocalized in space because their de Broglie lengths
are always larger than the inter-atomic distance,
and overlapping with each other.
As a result, instead of discrete atoms, quantum liquids 
must be described in terms of the fluid volume elements and elementary excitations 
such that non-locality and continuity are preserved
down to very short scales \cite{pi56,wyatt74,wyatt84}.
Historically, it was Landau \cite{lan41} who developed the theory of superfluidity
based on the mechanics of continuous media
which qualitatively agreed with experiment \cite{yabk,hw61,db98}.

However, some questions remain open in this regard - in particular, those
about the interplay between  original degrees of freedom ($^4$He atoms) 
and  emergent collective ones - such as the fluid volume elements
(also
called the fluid particles in the Lagrangian description \cite{yama}).
Indeed, the fluid-dynamical description of  liquid helium
presumes that  usage of the volume element's notion
must be physically justified before,
since it is the volume element, not the atom itself, which is supposed to be the
most elementary object of the fluid approach.
Thus, the understanding and describing of this notion in the quantum fluid theory
is more important than in the classical one because for classical fluids the continuity
is just the long-wavelength approximation whereas for  quantum ones it is an empirically
established fact. 
Therefore,
searches for the most suitable collective degrees of freedom in superfluids continue,
and
the issue of how  the formation and stability of the volume element of  liquid
helium as a continuous medium
can
be explained from first principles 
remains a subject of active studies nowadays \cite{kc01,ant03}.

The Bose-Einstein condensation (BEC)  is
another aspect of the superfluidity phenomenon
which must be taken into account \cite{yu11}.
Being predicted almost a century ago, it
was used by London and Tisza \cite{lo38}
to explain  the 
superfluidity
of liquid helium 
discovered by Kapitsa and Allen \cite{kaal38}.
The existence of BEC in the superfluid phase has been confirmed
by inelastic neutron scattering \cite{cw71,Dalfovo:1999zz},
thus, the influence of BEC upon the properties of liquid helium 
is under current study as well \cite{bs57b,lee57,ip11}.

Here we  describe 
the structure of the volume element of superfluid helium as well
as the phenomenological consequences which follow.
Among the derived effects is the famous Landau roton spectrum for which 
full theoretical explanation is still pending despite the numerous
efforts made towards its understanding \cite{fey53,ck09,ant03,kc01,bs57b,lee57,ip11}.
As a matter of fact, even usage of the term ``roton'' itself
is often a
purely historical one
since the explanation of the Landau spectrum in helium 
does not necessarily have
to be based on rotating degrees of freedom of helium atoms \cite{kc01,ip11}.
Indeed, here we show that it is possible to formulate
a theory without their explicit involvement - yet
obtain a remarkable
agreement with experiment.

Throughout the paper we neglect temperature effects which is a standard
assumption for fundamental models since below the lambda point
the thermal effects can be considered as corrections (the superfluid
helium has almost perfect heat conduction).

In section \ref{s-mod} we formulate a general theory of helium II,
in sections \ref{s-obs} and \ref{s-exp} we derive the
observable quantities
and compare them with available experimental data.
Conclusions are made in Sec. \ref{s-con}.

\scn{Collective variable theory}{s-mod}

We divide a theory into two nested parts.
The shorter-wavelength part
describes the interior structure of the volume element of helium superfluid
 using the non-linear logarithmic
quantum wave equation.
The longer-wavelength part
is the quantum many-body theory of these
elements, treated as new collective degrees of freedom.
While one assumes them to be effectively point-like objects in the long-wavelength approximation, their spatial 
extent and internal
structure are taken into account by virtue of the nonlocal interaction
term.
At that, these two models are not independent: in the shorter-wavelength part
we derive behavior and specific values of the parameters 
for the longer-scale one,
in accordance with the ``Russian-doll'' picture of nesting of scales.


As a starting point,
we assume that strongly-correlated helium atoms can form a bound
state characterized by a single macroscopical wavefunction.
The wave equation describing such object cannot be of the Gross-Pitaevskii
(cubic \schrod) type 
for at least two reasons.

The first one is that the
GP approach \cite{GP61}
is a perturbative one
which takes into account only two-body interactions
and neglects anomalous contributions to self-energy
which is a good approximation for dilute system like cold gases \cite{pethick04}, but
unlikely to suffice for liquids.
Indeed, according to aforesaid the atoms in  quantum liquid are
delocalized and thus nothing prevents them from getting involved into
multiple-body interactions
where the multiplicity can vary from two to the total number of atoms in
the system. 
An example of multi-body (three and more) interactions being very
important for forming bound states of bosons at low temperatures
is the Efimov state \cite{efim,nfj98}
which has been experimentally observed in helium \cite{efimexp},
a recent review can be found in \cite{kgm11}.

The second obstacle is that
the ground-state wavefunction of the free-space GP BEC model does not describe
a localized object.
Instead, the free GP condensate tends to occupy
all available volume - as such one needs to apply
an external potential trap
to confine the condensate and stabilize the system.
It is difficult to imagine that upon the transition into
a superfluid phase
the free helium suddenly becomes
surrounded, both in parts and as a whole, by a hypothetical external potential, 
not to mention
that the coherent appearance, stability and synchronization of such domains
across the 
bulk
would be impeded by significant volatility of the liquid. 
   
\begin{figure}[htbt]
\begin{center}\epsfig{figure=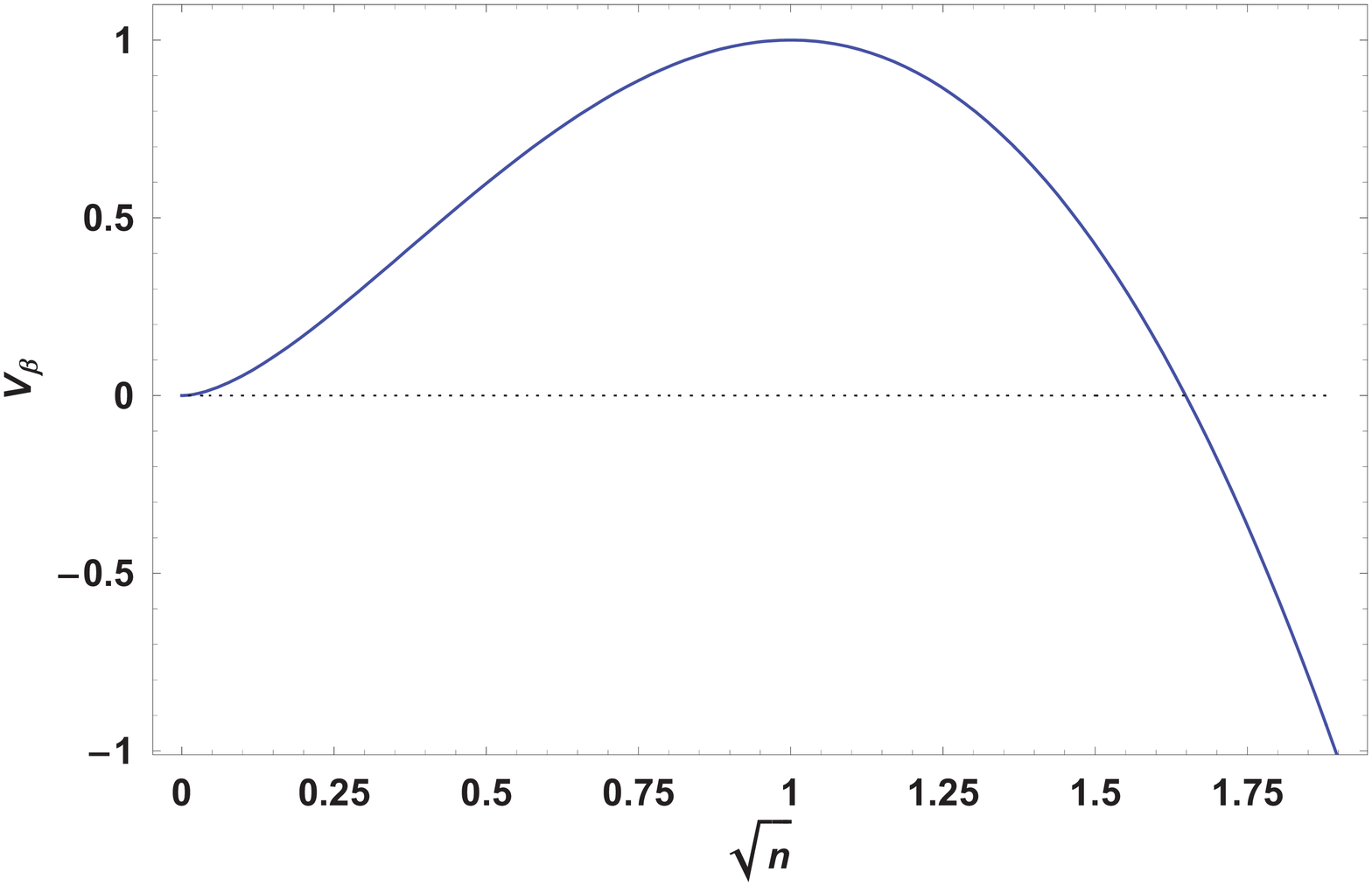,width=  1.01\columnwidth}\end{center}
\caption{
Field potential for the logarithmic BEC (in units $\beta \tilde a^3$) versus square root of particle
density 
(in units ${\tilde a}^{-3/2}$).
}
\label{f:ftpot}
\end{figure}

There exists, however, another candidate where these problems
simply do not occur in first place.
This is  non-linear Bose liquid defined by virtue of the logarithmic
\schrod~equation:
\be\lb{e-becgeneq}
\left[
- i \hbar \, \partial_t
- \frac{\hbar^2}{2 m} \vec \nabla^2
-
\beta^{-1} \ln{(\tilde a^3 |\Psi|^2)}
\right]
\Psi
= 0,
\ee
where 
$\Psi = \Psi (\vec x, t)$
is the wavefunction of condensate
normalized to the number of particles $ N$
-
such that
particle density
is determined as $n = |\Psi|^2$,
$m$ is the mass of the constituent particle
(helium-4 atom in our case, i.e., $m = m_{\text{He}} \approx 6.64 \times 10^{-24}$ g),
and
$\beta$ and $\tilde a$ are constant parameters of interaction.
The equation alone received attention a while ago 
as the simplest U(1)-symmetric wave equation
(apart from the conventional \schrod~one)
which satisfies the dilatation covariance and 
separability 
properties \cite{ros69,BialynickiBirula:1976zp}
but for the purposes of theory of quantum liquids 
it has been applied only recently  \cite{az11}.
While the second-quantized Hamiltonian which might lead
to such equation is not known yet, the equation itself
has some interesting properties and thus can be used
as a starting point.
The equation can be also derived from the field-theoretical
action with the Lagrangian 
\be
{\cal L}
 =
\frac{i\hbar}{2}(\Psi \partial_t\Psi^* - \Psi^*\partial_t\Psi)+
\frac{\hbar^2}{2 m}
|\vec\nabla \Psi|^2
+
V_\beta (|\Psi|^2)
,
\ee
where the potential is defined as
\be\lb{e-ftpot}
V_\beta (n) = -
\beta^{-1}
n
\left[
\ln{(n \tilde a^3)} -1
\right]
.
\ee
The latter opens down and has local non-zero maxima at $n_\text{ext} = \tilde a^{-3}$
for positive $\beta$, see Fig. \ref{f:ftpot}.
Despite the potential not being bounded from below as a function 
of $|\Psi|$, no density divergences arise since the wavefunction cannot take
arbitrarily large values, due to the normalization constraint.

It turns out that the logarithmic Bose liquid has a number of 
features suitable for our objectives:
it implicates not only binary but also multiple-body interactions
(when more than two bodies can 
scatter simultaneously),
and
its ground state is the so-called \textit{gausson} \cite{BialynickiBirula:1976zp} - a spherically-symmetric object  
which is localized 
and stable even in absence of a trapping potential,
with the interior  density obeying the Gaussian law
\be
n (\vec x)=
n(0)
\, \text{e}^{-(r/a)^2}
,
\lb{e-solnotrap}\ee
where
$n(0) = N /(\pi^{3/2} a^3)$
is the central particle density,
$r = |\vec x|$,
and 
$a = \hbar \sqrt{\beta/ 2 m}$
is the  characteristic radius.
In principle,
for practical purposes
the relation 
$
 a \approx \tilde a 
$
can be assumed throughout this paper, 
which essentially 
means that the parameters
of the short-scale model
are not entirely independent but must be bound by (at least) one
physical constraint,
$
\beta^{-1} \approx 
\tilde{p}^2/ 2 m
,
$ 
where
$\tilde{p} = \hbar / \tilde a$
is the de Broglie momentum corresponding to the length scale $\tilde a$.

It should be emphasized that the object
described by (\ref{e-solnotrap})
is different from the classical droplet since
it does not have border in a classical sense,
therefore, its stability is supported not by surface
tension but
by nonlinear quantum effects in the bulk \cite{az11}.

The long-wavelength part can be formulated as follows.
As long as we have established that the nucleation and stability of volume elements
of the Gaussian type
is justified on quantum-mechanical grounds,
it is natural to assume that below the critical temperature the atoms 
tend to form 
the fluid elements of this kind.
Therefore, new collective degrees of freedom arise, so
a theory must be formulated in terms of the volume elements rather than
of the helium atoms themselves \cite{ant03}.
Besides, as long as the bosonic liquid can no longer be assumed
homogeneous one cannot use all the results of section 4 from \cite{az11}.
Thus, we use the following Hamiltonian 
\be
\hat H=
\int d^3 x
\, 
\hat\psi^\dagger (\vec x)
\left(
-\frac{\hbar^2}{2 M} \vec \nabla^2
\right)
\hat\psi (\vec x)
+
\hat H_\text{int}
,
\ee
where 
$M$ is the mass of the volume element  and 
$\hat\psi$ is the corresponding field operator.
The interaction is defined via
the nonlocal term
\[
\hat H_\text{int}
=
\tfrac{1}{2}\!\int\!\int d^3 x\, d^3 x'
\, 
\hat\psi^\dagger (\vec x)
\hat\psi^\dagger (\vec x')
U\!
\left(
|\vec x - \vec x' |
\right)
\hat\psi (\vec x')
\hat\psi (\vec x)
,
\]
where 
$U (r)$ is the energy of interaction
between volume elements.
This energy can be estimated in the following way.
Using (\ref{e-solnotrap}) one can derive that the
Gaussian volume element of size $r \sim a$ stores an amount
of internal bulk mass-energy 
\[
\epsilon (r) \propto 
\int_0^r n (r') r'^2 dr' 
\propto
\frac{1}{a} (r - r_0) \, \text{e}^{-(r/a)^2}
\left[
1 + {\cal O} \left(r - a\right)
\right],
\]
where
the value
$r_0 = a
\left[
1/2 + 1/(\text{e} \sqrt{\pi}\, \text{erf}(1))
\right] \approx 0.75 \,a$ refers to the  point
where the dominant term of
$\epsilon (r)$ changes sign. 
Since each element has been shown to be
stable with respect to small
perturbations,
it tries to maintain its size and mass
when interacting with immediate environment,
therefore, to alter these values one has to supply the amount of energy which is proportional to $\epsilon (r)$.
In absence of strong external fields this energy 
can come only
via interaction with other volume elements 
hence we can conclude that
$U\!
\left(
|\vec x - \vec x' |
\right) \propto \epsilon\!
\left(
|\vec x - \vec x' |
\right)$.
Thus we can introduce the proportionality factor $U_0 (\vec x)$,
assume it to be a constant in a leading-order approximation,
and estimate the interparticle potential as:
\be\lb{e-intercl}
U (r) =
\frac{U_0}{a} (r - r_0) \, \text{e}^{-(r/a)^2}
,
\ee
up to the terms of order ${\cal O} \left(r-a\right)$ which are  assumed to be small.
These terms can be safely omitted unless
interaction deforms the elements so strongly that 
their interior structure 
cannot be
neglected anymore; but then the element's spherical symmetry
becomes deformed as well. 
The quantity $U_0 = - a U (0)/r_0 \approx - 1.34\, U (0)$
becomes the free parameter of the long-wavelength part of the theory.
If $U_0$ is positive then the critical radius $r_0$ determines
the  inter-element separation below which a pair of neighboring
volume elements becomes unstable against coalescence.
In principle, any possible effects of the deformations of elements
can be  accounted for by further upgrading the constant $U_0$
to a correspondingly derived function $U_0 (\vec x)$ but 
for the purposes of our current study the approximation
 $U_0 (\vec x) \approx U_0$
will do the job, as will be shown below.

\scn{Observables}{s-obs}

In this section we derive the analytical
expressions for the main observables
of the theory: energy of excitations,
structure factor and speed of sound.

\sscn{Energy of excitations}{s-excit}

To derive the energy spectrum of excitations one can use three alternative approaches.
The first one is based on the perturbation theory \cite{bs57b},
second is the standard Bogoliubov
transform \cite{Bogoliubov47},
and third is about analyzing small perturbations of the 
equation of motion for the field operator - as shown in \cite{ip11}
by the example of the semi-transparent sphere model.
In our case one can check that 
all three approaches yield the same spectrum in the leading approximation.
The Bogoliubov's method is the most straightforward and simple one, hence it
will be used from now on.
After some algebra
we arrive at the following $N$-body Hamiltonian operator
\be
\hat H \approx
\frac{1}{2}
n_0^2 \bar U_0 V
+ 
\sum_{\vec p \not=0}
E_p \hat a_p^\dagger \hat a_p
,
\ee
where 
$\hat a_p^\dagger $ and $\hat a_p $ are the creation and annihilation 
operators of the quasi-particle with momentum $\vec p$,
$n_0 = N/V = \rho_{\text{He}} / M$ is the background particle density of the liquid of $N$
elements occupying the volume
$V$, $ \rho_{\text{He}} \approx 0.145$ g/cm$^3$ is the liquid helium-4 density,
$\bar U_p = \int U (x) \, \text{e}^{i \vec p \cdot \vec x/\hbar} d^3 x $ is the Fourier transform of 
the interaction potential (\ref{e-intercl}),
and 
$E_p$ is the quasi-particle's energy
counted from a ground state.
The energy obeys the following
dispersion relation:
\be\lb{e-disp}
E_p=
\frac{p^2}{2 M}
\sqrt{1 + \frac{4 n_0 M \bar U_p}{p^2}}
,
\ee
with the Fourier transform being calculated as
\be\lb{e-up}
\bar U_p
=
\pi a^3 U_0
\left(
1- \sqrt{\pi} f_k
\, \text{e}^{-(a k/2)^2}
\right)
,
\ee
where
$
f_k
=
r_0/a
+
\left[
\tfrac{1}{2}
a k
-
(a k)^{-1}
\right]
\text{erfi} (a k/2)
$
and
$\vec k = \vec p/\hbar$ is the wave vector.

\begin{figure}[htbt]
\begin{center}\epsfig{figure=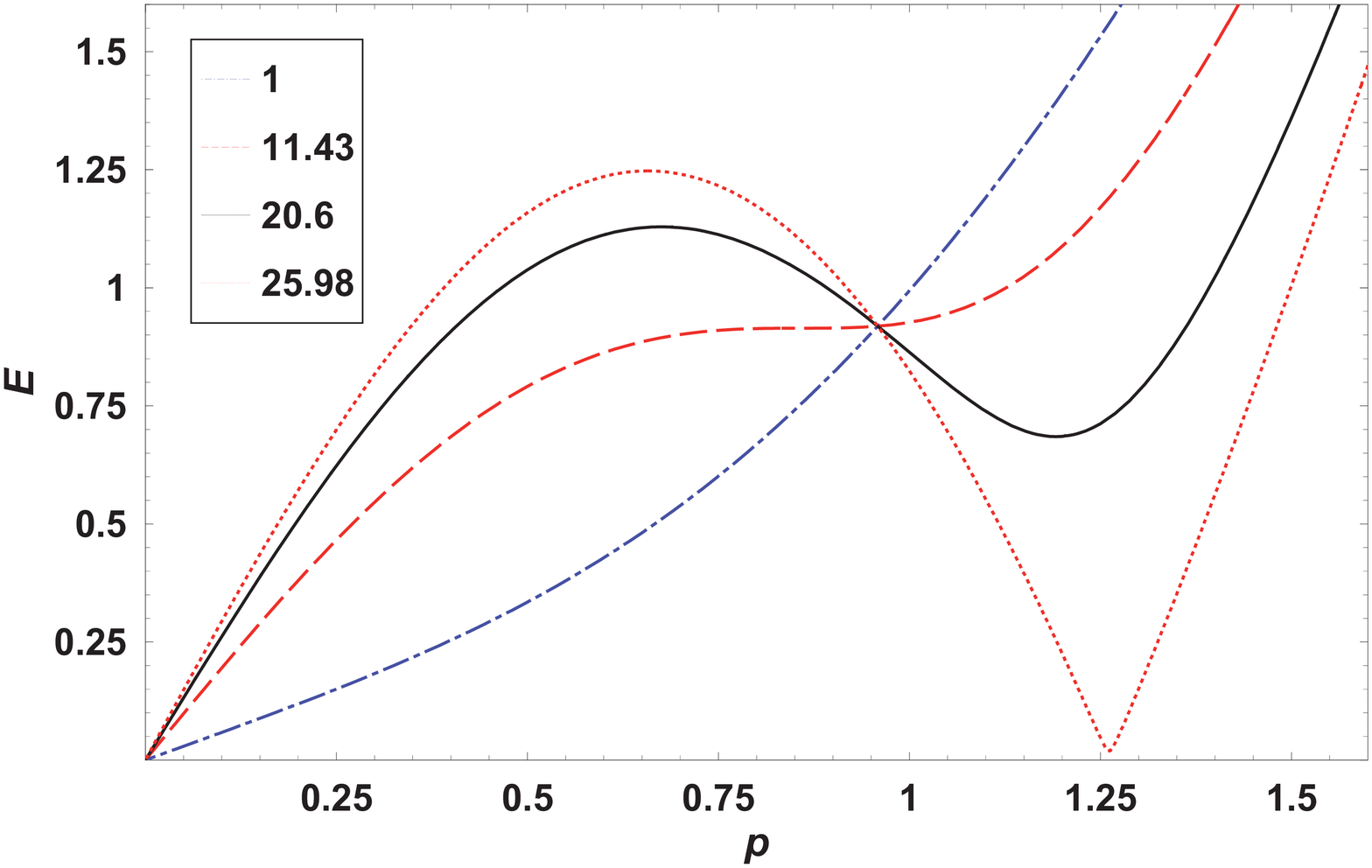,width=  1.01\columnwidth}\end{center}
\caption{Energy of quasi-particles $E_p$ versus momentum (in units of $E_a$ and $p_a$, respectively)
for different values of  $ u_0 $.
Below the value $11.43$ the roton minimum disappears,
and  above  $25.98$ energy becomes complex-valued.}
\label{f:allu1}
\end{figure}

\begin{figure}[htbt]
\begin{center}\epsfig{figure=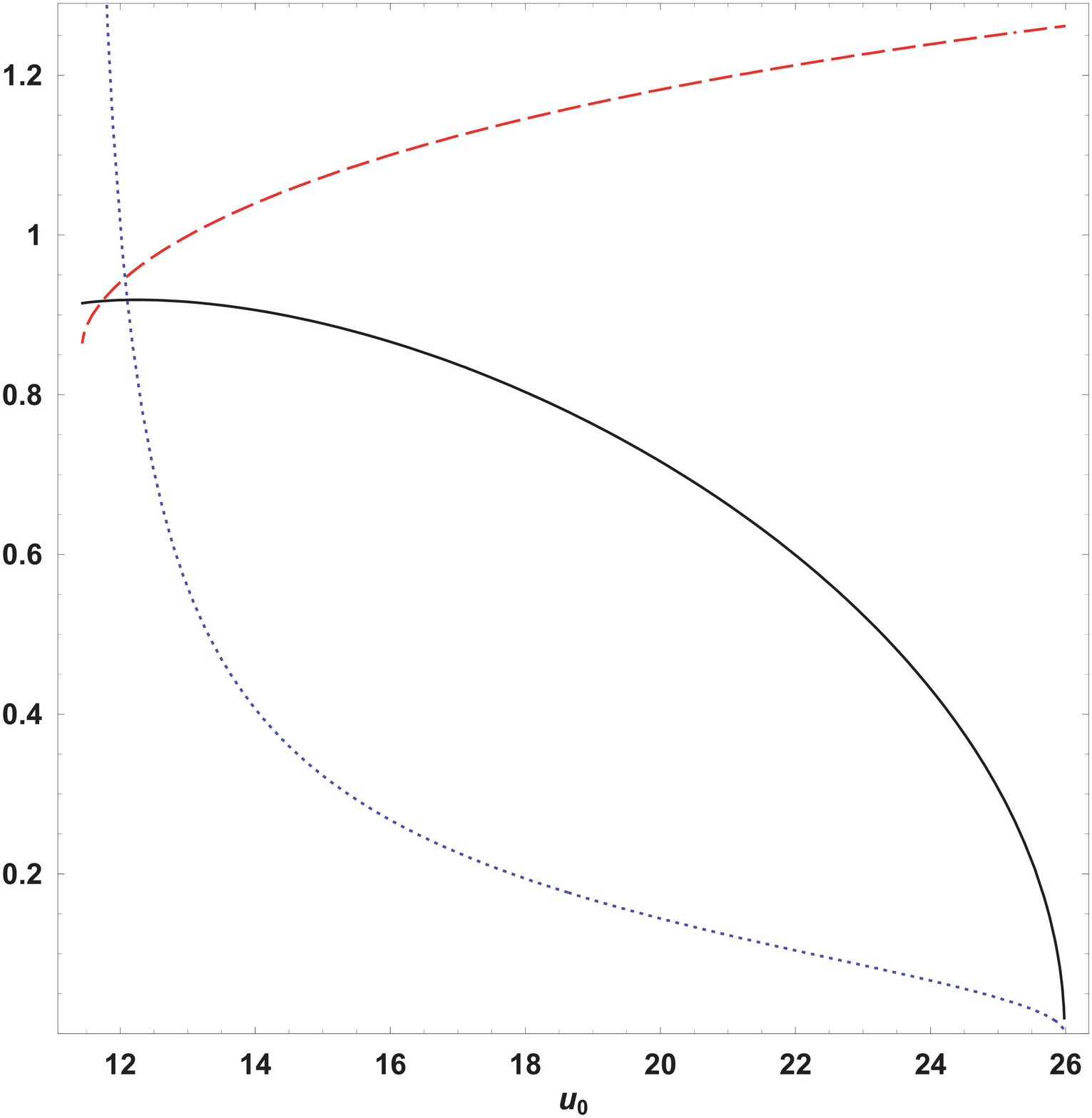,width=  0.9\columnwidth}\end{center}
\caption{Local minimum's values of momentum $p_0/p_a$ (dashed curve), roton energy gap $\Delta/E_a$
(solid)
and mass $\mu/M$ (dotted)
versus the interaction coupling constant
taken between $u_0 \approx 11.43$ and $25.98$.}
\label{f:minallu}
\end{figure}

Introducing
the volume element's de Broglie momentum scale
$p_a = 2 \hbar/a$
and corresponding energy scale $E_a = p_a^2 / 2 M$,
one can re-write the dispersion relation (\ref{e-disp}) in the dimensionless
form
which shows that the dynamics of the quasi-particle
depends on just one essential (non-scale) parameter 
\be
u_0 = 2\pi n_0 a^5 M U_0/\hbar^2
.
\ee
Numerical analysis shows that admissible
values of this parameter lie 
between $11.43$ and $25.98$ where the local (so-called roton)
minimum appears, see Fig. \ref{f:allu1}.
In its vicinity 
the dispersion relation can be written in the Landau form,
$
E_p \approx \Delta + \frac{(p - p_0)^2}{2\mu}
,
$
where $\Delta $ is the roton energy gap,
$p_0$ is the minimum value of momentum,
$\mu $ is the effective roton  mass.
The corresponding profiles for different values of $u_0$
are given in Fig. \ref{f:minallu}
from
where one can see that
the roton mass diverges at the lower bound (the local minimum becomes a saddle point)
whereas the energy becomes complex-valued above the upper bound.

\begin{figure}[htbt]
\begin{center}\epsfig{figure=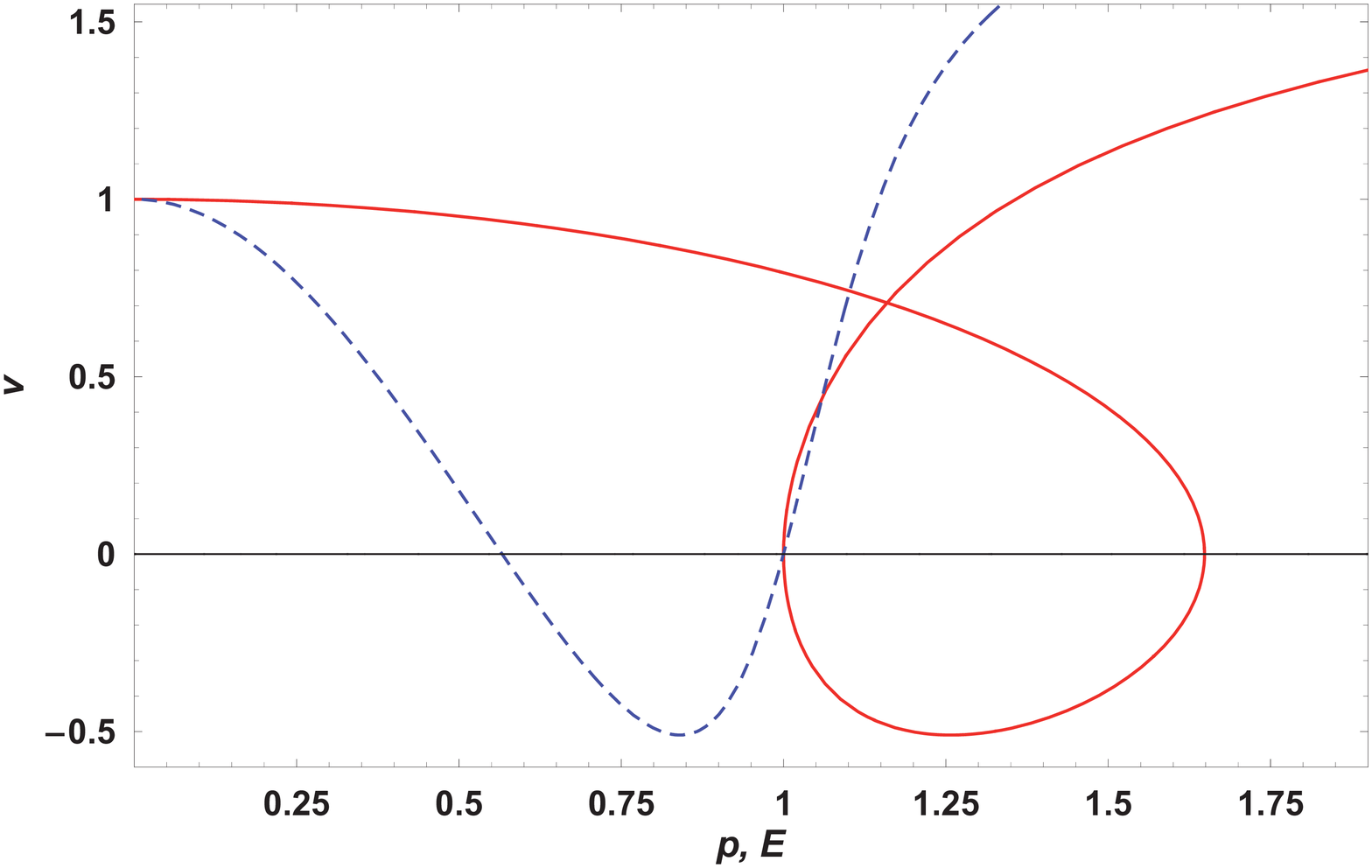,width=  1.01\columnwidth}\end{center}
\caption{Quasi-particle's velocity $v$ (in units of $v|_{p=0}$) versus momentum (in units of $p_0$, dashed curve)
and energy (in units of $\Delta$, solid curve).
It is plotted at $u_0 \approx 20.6 $ (then $v|_{p=0} \approx 273$ m/s) but
such behavior holds qualitatively for $u_0$ between $11.43$ and $25.98$.
}
\label{f:velo}
\end{figure}

We can also plot the 
quasi-particle's velocity
$\vec v = 
v \vec\ell$
where
$v = 
\partial E_p / \partial p$
and $\vec \ell = \vec p / p$.
From Fig. \ref{f:velo} one can see that within the above-mentioned range of
the parameter $u_0$ the liquid indeed exhibits 
the superfluidity feature - 
an interval of momentum
and energy 
over which 
the effective mass 
$p/v$ 
turns negative
always exists,
thus
indicating that the production of dissipative excitations is 
suppressed.
As a matter of fact, this is just a restatement of the original Landau idea of
introducing the roton minimum into the excitation spectrum to explain
the superfluidity - since a negative $p/v$ range always appears if a positive-definite 
function has the local minimum preceded by a local maximum.
One can also notice that at small momenta (phonon regime) the velocity varies very slowly
as a function of frequency $\omega = E_p / \hbar$, as if there is no dispersion at all.

\sscn{Speed of sound}{s-snd}

By sound here we  understand 
the conventional acoustic oscillations,
the heat transfer and second sound are not considered since
the temperature effects are neglected throughout the paper, as mentioned in the introduction.

Thus, our observable of interest is a mean value of the sound velocity
$\langle v_s \rangle$ which
can be estimated as follows.
If size and interior structure of volume elements are neglected
then the speed of sound would be given just by
the standard formula \cite{pethick04}
\be
v_s^{(0)} = \sqrt{n_0 \bar U_0/M} = \lim\limits_{p\to 0} v .
\ee
However, in general one should take into 
account that sound propagation is affected by the volume element's ``interior''
which is described by the logarithmic Bose liquid  in a ground state,
and
speed of sound inside the latter is known to be
\be
c_s = 1/\sqrt{m |\beta |}
,
\ee
in the leading
approximation \cite{az11}.
The mean speed of sound thus can be estimated as an averaged
sum of the background and excitation contributions taken with the weights 
controlled by the interior  density of a volume element (\ref{e-solnotrap}),
\bw
\be\lb{e-ssnd}
\langle v_s \rangle
=
{\bar a}^{-1} \int\limits_0^{\bar a}
\left[
  \chi c_s +
\left(
1 - 
\chi
\right) v_s^{(0)}
\right] d r
=
v_s^{(0)}\!
\left[
1
-
\frac{\sqrt{\pi}}{4}
\frac{a}{\bar a}
 \,
\text{erf} (\bar a/a)
\left(
1 - \frac{c_s}{v_s^{(0)}}
\right)
\right]
,
\ee
\ew
where 
$\chi = (1/2)\, n(\vec x)/n(0) = \tfrac{1}{2} \text{e}^{-(r/a)^2}$ is the weight factor,
$\tfrac{4}{3}  \pi \bar a^3 = 1/n_0 = M/\rho_{\text{He}} $
is the effective packing volume of an element.
The numeric coefficient inside the weight factor has been chosen in such a way as to satisfy
the following limit case properties:
if $a\to 0$ then the $v_s^{(0)}$ term dominates (zero-size limit),
and if $a\to \infty$ then both terms contribute equally.

\sscn{Structure factor}{s-sfa}

According to (\ref{e-solnotrap}), our volume elements are essentially
quantum states of delocalized atoms which do not have a border
or interface
in a 
classical sense.
Therefore, their presence in the liquid helium can be judged
only indirectly - via the structure factor $S_k$ which
determines the scattering of
neutrons (or x-rays, after multiplication by the atomic
structure factor) and can be measured
experimentally. 
This situation, however, is not something extraordinary:
the microscopical objects employed in
previous models of liquid helium, such as 
vortex rings \cite{fey53}, hard spheres \cite{lee57,ip11} or stochastic clusters \cite{kc01},
were never claimed to be ``directly seen'' (whatever it could mean in our essentially quantum context)
either.
Instead, if a selected model's predictions,
e.g. for the excitation energy  and structure factor, 
coincide with experimentally observed values
then the corresponding microscopical interpretation can be adopted - at least,
until the better
one arrives.

The computation of the structure factor is slightly more difficult
than that of the  energy of excitations: as shown in the previous 
section, the computation of $E_p$ requires, basically, the knowledge
of the inter-element interaction potential, such as (\ref{e-intercl}), only.
By contrast, to derive the structure factor one should
derive the excited-state wavefunction of the liquid helium 
which can be done only approximately under certain simplifying assumptions.
If one considers the ``bare'' helium atoms to be the
fundamental degrees of freedom 
then the structure factor
would be given by the formula derived in \cite{fey53}.
In our case it would yield
\be\lb{e-sfold}
S_k^{(0)}
=
p^2 / 2 M E_p
=
\left(
1 + \frac{4 n_0 M \bar U_p}{p^2}
\right)^{-1/2}
,
\ee
where all quantities have been defined in the previous section. 
However, this formula
can not be blindly applied to our case
since we use the collective degrees of freedom which are not 
``bare'' helium atoms but their bound states.
Besides, the logarithmic non-linearity
affects not only the energy of excitations but
also the correlation functions themselves.
Therefore, one needs to slightly revise the Feynman's derivation
by
taking these effects into account - especially if one considers that
Feynman and Cohen's theory gives only a qualitative
agreement with experiment. 
Fortunately, the computation is straightforward
and necessary corrections are minimal (yet important).
We begin by assuming that the
total energy of the system of $N$ fluid elements comes
from minimizing the integral
$
{\cal E} =
\int
\psi^* {\cal H} \psi \, d^{3 N}\!x
, 
$
subject to the fixed normalization integral
$
{\cal J} =
\int
\psi^* \psi \, d^{3 N}\!x
, 
$
where $\psi$ is the excited-state wavefunction.
Throughout this section we use
$\vec x^N$ to denote the set of coordinates $\vec x_i$ of all
the elements, and $\int d^{3 N}\!x$ to represent the integral over all of them.
The Hamiltonian of the system is given by a standard
expression,
\be
{\cal H} 
=
{\cal H}_0 + V
=
- \frac{\hbar^2}{2 M}
\sum\limits_i \vec\nabla_i^2 
+ V 
,
\ee
where $V$ is the  potential energy of the system
and Latin indices run from 1 to $N$.
Following Feynman,
for the excited-state wavefunction we assume the Bijl ansatz \cite{bi40}:
\be\lb{e-bijl}
\psi = F \psi_0 
=
\sum\limits_i f (\vec x_i) \psi_0
,\ \
f (\vec x) \equiv \text{e}^{ i \vec k \cdot \vec x},
\ee
with the only difference that
$\psi_0 = \psi_0 (\vec x^N)$ is now the ground-state wavefunction of 
one-particle states of logarithmic quantum liquid
(not plain atoms as was in the original Feynman derivation).
Due to the separability property of the logarithmic \schrod~equation,
they
satisfy
its stationary many-body analogue: 
\be\lb{e-nblse}
{\cal H}_0 \psi_0 = 
\frac{m}{M}
\left[
N E_0
+
\beta^{-1} 
\ln{(\tilde a^{3 N} |\psi_0|^2)} 
\right]
\psi_0
,
\ee
where
the logarithmic term has replaced the potential part, 
according to the two-scale structure of our theory
(we remind also that throughout the
paper the external potential is neglected).
With the use of the separability property 
the $N$-particle equation (\ref{e-nblse})
can be  
analytically solved for the case of a ground state,
cf. section \ref{s-mod}.
One thus
obtains  the 
eigenvalue  
$
E_0 = 
 3 \beta^{-1}
\left[ 1+ \ln{(\sqrt{\pi} a /\tilde a)} \right]
\approx
 3
\beta^{-1}
\left(
1+
\frac{1}{2}
\ln{\pi}
\right)
$
and
$
\psi_0 \propto 
\prod\limits_i \Psi (\vec x_i)
\propto 
\prod\limits_i
\exp{(- |\vec x_i|^2/2 a^2)}
$,
cf.
\cite{BialynickiBirula:1976zp}.
 
Further, with the ansatz (\ref{e-bijl}) in hand the energy per particle
of the system 
can be approximately written as
\bw
\be
E_p + E_0 =
\frac{
\int
\left[
\frac{\hbar^2}{2 M}
\frac{1}{N}
\sum\limits_i
\vec\nabla_i F^*
\cdot
\vec\nabla_i F
+
\frac{m}{M}
|F|^2
\left(
E_0
+
\beta^{-1}
 \ln{(\tilde a^3  \rho_N^{1/N})}
\right)
\right]
\rho_N
d^{3 N}\!x
}{
\int
|F|^2 
\rho_N
d^{3 N}\!x
}
,
\ee
where 
\be
\rho_N 
= 
\left(
\frac{1}{\pi^{3/2} a^3}
\right)^N
\text{e}^{- \sum\limits_i |\vec x_i|^2/a^2}
\ee
is the ground-state density function.
By construction, excitation energy $E_p$ 
takes into account also
the inter-element interaction, see Sec.~\ref{s-excit}.
Further, using the fact that logarithm varies significantly slower
than the Gaussian,
we obtain 
\be
E_p + E_0
 - 
\frac{m }{M}
\left[
E_0
+
\beta^{-1}
\ln{(\tilde a^3  \rho_N^{1/N} (0))}
\right]
=
\frac{\hbar^2}{2 M}
\frac{
\frac{1}{N}
\sum\limits_i
\int
\vec\nabla_i F^*
\cdot
\vec\nabla_i F
\rho_N
d^{3 N}\!x
}{
\int
|F|^2 
\rho_N
d^{3 N}\!x
}
,
\ee
which after integration reduces to
\be 
E_p 
+ 
\Sigma
=
\frac{\hbar^2}{2 M}
\frac{
\int
\vec\nabla f^* (\vec x)
\cdot
\vec\nabla f (\vec x)
d^3 x
}{
\int
f^* (\vec x_1)
f (\vec x_2)
P (\vec x_1 -\vec x_2)
d^3 x_1 d^3 x_2
}
,
\ee
where
$P (\vec x_1 -\vec x_2)$ is the probability of finding an element at $\vec x_2$ per unit volume
if one is known to be at $\vec x_1$,
and we denoted
\[
\Sigma
\approx
E_0
-
\frac{m}{\beta M}
\left(
3+
\ln{\pi}
\right)
\approx
3
\beta^{-1}
\left[
1 - \frac{m}{M}
+
\frac{1}{2}
\left(1 - \frac{2}{3}
\frac{m}{M}
\right)
\ln{\pi}
\right]
.
\]
\ew
The variation with respect to $f^*$
yields  the equation
\be
\left(
E_p + \Sigma
\right) 
\int
f (\vec x_2)
P (\vec x_1 -\vec x_2)
d^3 x_2
+
\frac{\hbar^2}{2 M}
\vec\nabla^2 f (\vec x_1)
=
0
,
\ee
and upon remembering (\ref{e-bijl}) this gives us
the final formula for the structure factor
$
S_k 
\equiv
\int P (\vec x) \text{e}^{ i \vec k \cdot \vec x} d^3 x
$
in a logarithmic theory:
\be\lb{e-sf}
S_k
=
\frac{p^2}{2 M \left(
E_p + 
\Sigma
\right) }
=
\left[
\sqrt{1 + \frac{4 n_0 M \bar U_p}{p^2}}
+
\frac{2 M
\Sigma }{ p^2}
\right]^{-1}
.
\ee
Thus, the only difference from the ``non-logarithmic'' formula (\ref{e-sfold})
is the appearance of an additive constant term in the denominator 
which is induced by the logarithmic nonlinearity.

\scn{Theory versus experiment}{s-exp}

\begin{figure}[htbt]
\begin{center}\epsfig{figure=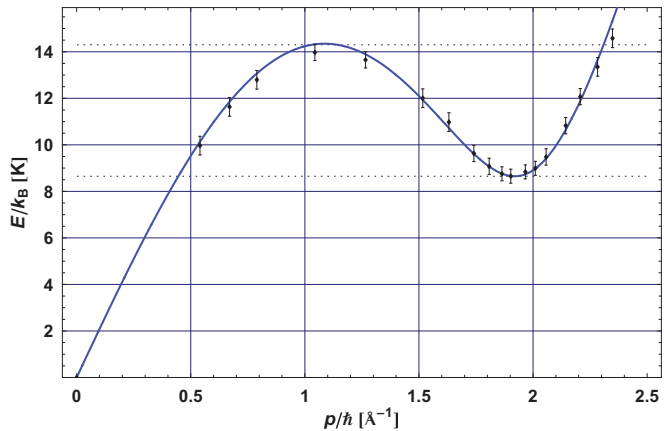,width=  1.01\columnwidth}\end{center}
\caption{Energy of quasi-particles $E_p$
versus wave vector.
Solid curve is a theoretical prediction,
dots denote experimental data \cite{yabk}.}
\label{f:experim}
\end{figure}

In this section we compare the derived observables
of our model with available experimental data.
If we choose the value $u_0 \approx 20.6$ then the established experimental data for
energy of excitations (see, for instance \cite{yabk}) can be 
successfully fit.
When fitting, we do not use the least-squares techniques
but rather aim at the precise position of the roton minimum -
so as to estimate the magnitude of deviations from experiment at the neighboring local  maximum $E_\text{max}$
at $p = p_\text{max}$ also known as the \textit{maxon} peak.
It turns out that 
our theory can reproduce the maximum with accuracy which is 
better than three percent, see
the cumulative 
comparison between theory and experiment
given in Fig.\ref{f:experim} and Table \ref{t:tabex}.
The other values of primary and secondary parameters of the 
theory computed 
at $u_0 \approx 20.6$
turn out to be
the following:
\ba
&&
\bar a  \approx 2.37\ \text{\AA},\
M / m \approx 1.228,\
U_0 / \Delta \approx 69,\
\nn\\&&
a \approx 1.25 \ \text{\AA},\
p_a/p_0 \approx 0.84,\
E_a/\Delta \approx 1.46,\
\lb{e-fitted}\\&&
\beta^{-1}/k_B \approx 3.84 \ \text{K},\
\beta^{-1}/E_a \approx 0.31
,
\nn
\ea 
hence this set 
will be used for further comparisons with 
experiments in this paper.
For instance, using (\ref{e-fitted}) the velocity-related quantities from section \ref{s-snd}
are evaluated as
$c_s \approx 89$ m/s and $v_s^{(0)}\approx 273$ m/s, therefore,
the mean value of the speed of sound (\ref{e-ssnd})
can be estimated as
\be
\langle v_s \rangle \approx 231 \ \text{m/s}
,
\ee
which is in a good agreement with experiment as well \cite{ps47}.

\begin{figure}[htbt]
\begin{center}\epsfig{figure=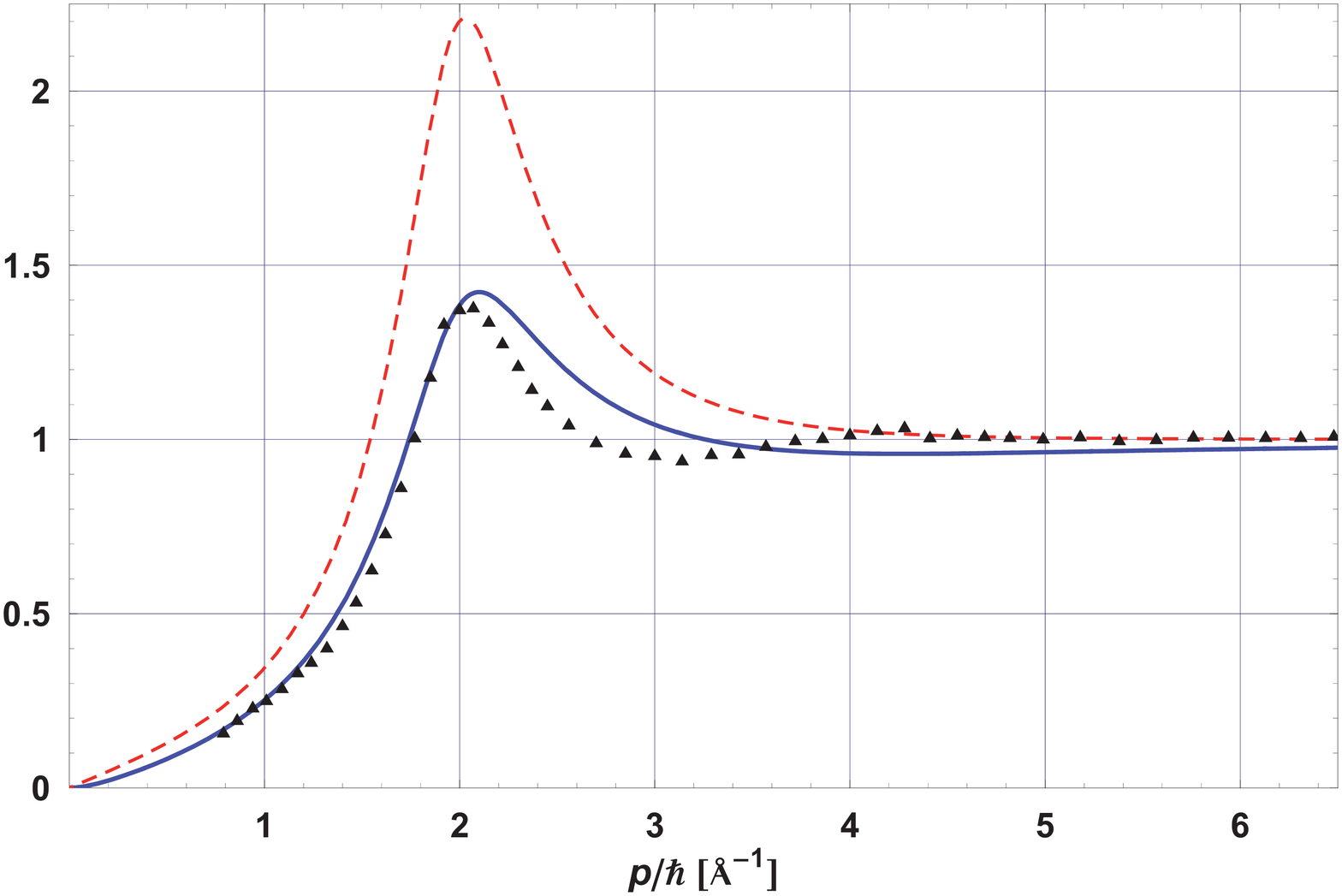,width=  1.01\columnwidth}\end{center}
\caption{Structure factor $S_k$
versus wave vector.
Solid curve is a theoretical prediction given by (\ref{e-sf}) evaluated at the values (\ref{e-fitted}),
triangles denote the experimental data \cite{db98}. 
For the sake of comparison, the result
of the formula (\ref{e-sfold})
is being plotted
as well (dashed curve).
}
\label{f:expsf}
\end{figure}

Finally, 
the structure factor (\ref{e-sf}) is also evaluated using 
the values (\ref{e-fitted}).
We compare the theoretical curve with the 
neutron scattering data in Fig. \ref{f:expsf}
(the neutron scattering is the most relevant method here
since it is the one which is being used for measuring $E_p$ as well).
One can see that the asymptotic behaviour of the structure factor, both at small 
and large momenta, is compatible with known experimental data,
see also the discussion in the concluding section. 
For instance, from the scattering data one can not surely conclude that 
at small momenta $S_k$ 
must tend to zero as a linear function of $p$.
Indeed, as one can see from 
\cite{db98} (see also the first figure from the Feynman-Cohen paper in \cite{fey53}),
$S_k$ is not a straight line near origin but rather a curve which
resembles more of a parabola and thus agrees with the asymptotic behaviour  
of (\ref{e-sf}) at $p \to 0$. 
Though, one can not {\it a priori} exclude that in real scattering
experiments some side effects might appear and introduce
the asymptotic corrections proportional to $p$ and
thus override
the non-linear asymptotics; however,
these additional effects are not a subject of our study at 
this stage.

\begin{table}[b]
\caption{\label{t:tabex}%
Theory versus experiment}
\begin{ruledtabular}
\begin{tabular}{llll}
\textrm{Quantity}&
\textrm{Theory}&
\textrm{Experiment}
&
\textrm{Difference,\%}\\
\colrule
$\Delta/k_B$, K & 8.65 (fit) & 8.65$\pm$0.04 & 0$\pm$0.4\\
$p_0/\hbar $, \AA$^{-1}$ & 1.91 (fit)  & 1.91$\pm$0.01 & 0$\pm$0.5\\
$\mu / m$ & 0.16 & 0.16$\pm$0.01 & 0$\pm$6\\
$E_\text{max}/k_B$, K  & 14.3 & 13.92$\pm$0.1 & 2.7$\mp$0.7\\
$p_\text{max}/\hbar $, \AA$^{-1}$ & 1.08  & 1.11$\pm$0.04  & 2.6$\pm$3.6\\
$\langle v_s \rangle $, m/s & 231  & 237$\pm$2 & 2.6$\pm$0.08\\
\end{tabular}
\end{ruledtabular}
\end{table}

\scn{Conclusion}{s-con}

The superfluidity of liquid helium II is the complex phenomenon 
exhibiting a unique feature of the Landau spectrum of excitations which
successfully explains the behavior of specific heat, viscosity and sound velocity. 
The microscopical theory of this phenomenon must address many 
different aspects such as the Bose-Einstein condensation, appearance of collective degrees of 
freedom, non-locality and continuity, formation of the volume element of the liquid, its structure and stability. 

Here we proposed a concise analytical theory of structure and excitations in superfluid helium which elaborates on these aspects.
It consists of two models which act on different length scales, but
are connected via the parametric space: the behaviour of quantities
and the values of parameters in the long-wavelength model are derived from the 
short-wavelength part.
The latter advocates the appearance of the collective
degrees of freedom which justify the possibility of the fluid-dynamical description and
can be used for characterize the volume elements
of the quantum liquid. 
The ``interior'' structure of a fluid volume element is described 
using the non-perturbative approach based on the logarithmic wave equation.
It turns out that the interior density of the element
obeys the Gaussian distribution. 

The long-wavelength part is the quantum many-body theory of the volume elements
as effectively point-like objects - yet their spatial 
extent and internal
structure are taken into account by virtue of the nonlocal interaction
term. 
The corresponding interparticle interaction potential (\ref{e-intercl}) was not postulated but 
derived from the short-wavelength part,
and it 
appeared to be very simple as compared to the empirical and semi-empirical inter-molecular 
potentials widely used in perturbative approaches \cite{ck09}.
On the other hand, more fundamental approaches, such as those
based on 
vortex rings \cite{fey53}, hard spheres \cite{lee57,ip11} or stochastic clusters \cite{kc01},
either
have shown so far only the qualitative agreement with experiment,
to our best knowledge, or some of observables in those models were not computed.
In our theory the quantitative agreement
with various experimental data
has been achieved:
with only one essential parameter in hand, $u_0$, we reproduced at high accuracy
(better than three per cent)
not only the roton minimum but also the neighboring local (maxon) maximum.
The velocity of sound and structure factor are also computed and found to be in a very good agreement
with experiment.
Slight deviations of the energy spectrum and
structure factor curves from experimentally observed ones
at large values of momenta $p \gtrsim 2\hbar $ \AA$^{-1}$
come due to 
the scattering particles begin to probe the interior of fluid elements
and thus
affect the corresponding wavefunctions.
At even higher momenta, $p \gtrsim 4\hbar $ \AA$^{-1}$, the agreement will eventually recover since the contributions from
kinetic energy come to predominate those from interaction.  

To conclude, the collective variables we have been using ultimately provide a unified description of the phonon, maxon and roton excitations. 
In future it would be interesting to study the applications 
of the models with the non-polynomial ground-state wave equations 
and Gaussian-like interparticle potentials, like
(\ref{e-becgeneq}) and (\ref{e-intercl}),
in the physical situations when superfluid becomes effectively low-dimensional (cigar- or disk-shaped)
and/or
subjected to external fields or admixtures.
A parallel direction of research would be to take into account temperature effects (see the
introductory section),
such as heat transfer, second sound, etc., which is necessary
for a number of practical applications.

\begin{acknowledgments}
I am grateful to O. Zaslavskii for organizing my visit to Karazin Kharkov National University
and Institute of Low-Temperature Physics during which main ideas of this paper nucleated,
as well as to A. Sergi for organizing my visit to the University of Messina.
Useful comments from A. Avdeenkov, S. Dolya and O. Marago as well as the proofreading
of the manuscript by P. Stannard are greatly acknowledged.
This work is based upon research supported by the National Research Foundation of South Africa.
\end{acknowledgments}


\def\AnP{Ann. Phys.}
\def\APP{Acta Phys. Polon.}
\def\CJP{Czech. J. Phys.}
\def\CMPh{Commun. Math. Phys.}
\def\CQG {Class. Quantum Grav.}
\def\EPL  {Europhys. Lett.}
\def\IJMP  {Int. J. Mod. Phys.}
\def\JMP{J. Math. Phys.}
\def\JPh{J. Phys.}
\def\FP{Fortschr. Phys.}
\def\GRG {Gen. Relativ. Gravit.}
\def\GC {Gravit. Cosmol.}
\def\LMPh {Lett. Math. Phys.}
\def\MPL  {Mod. Phys. Lett.}
\def\Nat {Nature}
\def\NCim {Nuovo Cimento}
\def\NPh  {Nucl. Phys.}
\def\PhE  {Phys.Essays}
\def\PhL  {Phys. Lett.}
\def\PhR  {Phys. Rev.}
\def\PhRL {Phys. Rev. Lett.}
\def\PhRp {Phys. Rept.}
\def\RMP  {Rev. Mod. Phys.}
\def\TMF {Teor. Mat. Fiz.}
\def\prp {report}
\def\Prp {Report}

\def\jn#1#2#3#4#5{{#1}{#2} {\bf #3}, {#4} {(#5)}} 

\def\boo#1#2#3#4#5{{\it #1} ({#2}, {#3}, {#4}){#5}}




\end{document}